\documentclass[]{aastex631}
\usepackage{amsmath}
\usepackage{txfonts}
\usepackage{graphicx}
\usepackage{comment}
\usepackage{hyperref}
\usepackage{multirow}
\usepackage{booktabs}
\usepackage{tabularx}

\begin{document}

\title{Analytical Constraints on the Radius and Bulk Lorentz Factor in the Lepto-Hadronic One-Zone Model of BL Lacs}

\author{Zhi-Peng Ma }
\affiliation{Department of Astronomy, School of Physics, Huazhong University of Science and Technology, Wuhan 430074,
 China}

\author[0000-0003-4976-4098]{ Kai Wang}
\affiliation{Department of Astronomy, School of Physics, Huazhong University of Science and Technology, Wuhan 430074,
 China}

\begin{abstract}

In this work, we study the parameter space of neutrino-emitting BL Lacs under the framework of the one-zone lepto-hadronic model. We show that constraints on the model come from various aspects of observations such as the variability timescale of blazar flares, gamma-ray opacity and the spectral energy distribution of electromagnetic emission, as well as the inferred neutrino emissivity of the blazar. We apply our method to two potential neutrino sources, i.e., TXS 0506+056 and PKS 0735+178, which are BL Lacs. Then, we explore and summarize the allowed range of parameters such as the bulk Lorentz factor and the blob radius under different distributions of injected protons. We find that the parameter space that is available to explain the BL Lac--neutrino association is sensitive to the proton distribution, and usually, an injected proton luminosity significantly exceeding the Eddington luminosity is required for both sources. Our results suggest that the simple lepto-hadronic one-zone model may not be a reasonable interpretation for BL Lac--neutrino associations.

\end{abstract}


\section{Introduction} \label{sec:intro}

In  September 2017, a~high-energy neutrino event, IceCube-170922A, with~the energy of $\sim$290 TeV was detected by the IceCube Observatory~\cite{telescope2018multimessenger}. With~good angular resolution, it had a 3$\sigma$ chance correlation with object TXS 0506+056, a~BL Lac object which was in a gamma-ray flaring state. Extensive efforts have been made to understand their ~\citep{ansoldi2018blazar,keivani2018multimessenger,padovani2018dissecting,sahakyan2018lepto,zhang2018variability,banik2019describing,cerruti2019leptohadronic,laha2019constraints,liu2019hadronuclear,padovani2019txs,righi2019neutrino,xue2019two,cao2020self}. The~event IceCube-170922A (hereafter IC170922A) and the following electromagnetic observations, for~the first time, directly indicate that blazars are potential sources of high-energy neutrinos~\citep{telescope2018multimessenger}.

Blazars are the most extreme form of active galactic nuclei (AGN), which have their jet pointing to observers approximately~\citep{urry1995unified}. As~one of the most powerful astrophysical persistent objects, blazars are widely considered as source candidates for the origin of extragalactic high-energy cosmic rays and neutrinos~\citep{mannheim1992neutrinos,atoyan2001high,murase2014diffuse,padovani2015simplified,petropoulou2015photohadronic,padovani2016extreme}. They are sub-classified as flat spectrum radio quasars (FSRQs) and BL Lacs objects depending on differences between their optical emission line features~\citep{urry1995unified}. The most significant characteristic of the spectral energy distributions (SEDs) of such objects is the two-hump structure, with different interpretations in various models. In~leptonic models, the low-energy hump of the SED is considered as a result of the synchrotron radiation of relativistic electrons in the jet, while the high-energy hump originates from inverse Compton (IC) scattering between high-energy electrons and low-energy photons from an external photon field (external Compton, EC)~\citep{rodrigues2019leptohadronic} or the electron-synchrotron radiation field (synchrotron-self Compton, SSC)~\citep{dermer1992high,bloom1996analysis,maraschi1992jet}. Such conventional models achieve great success in explaining the SED of blazars in the literature~\citep{mastichiadis1996variability} but fail to explain neutrino emission because of the lack of hadronic processes. Hence,~hadronic processes, i.e., the~photomeson process ($p\gamma$) or the proton-proton collision process ($pp$), have to be involved to 
be responsible for the neutrino emissions of blazars by considering the accelerated proton component~\citep{dar1997hadronic,araudo2013gamma,petropoulou2015photohadronic,ansoldi2018blazar,keivani2018multimessenger,cerruti2019leptohadronic,wang2018jet,zhang2020neutral}. In~these models, the low-energy bump of the SED is still dominated by the synchrotron radiation of accelerated electrons, as is the same as for leptonic models, while the high-energy hump could be from the superposition of the EC, the~SSC,~proton synchrotron radiation and/or the cascade emission of secondaries of hadronic processes~\citep{gasparyan2021time}.

Based on the number of emission regions, the~theoretical models can be classified as a one-zone model and a two-zone (or multi-zone) model \citep{gao1807interpretation,murase2018blazar,ansoldi2018blazar,keivani2018multimessenger,cerruti2019leptohadronic,liu2019hadronuclear,xue2019two,xue2021two}. More parameters are  invoked in the two-zone model to explain the multi-messenger observations (electromagnetic radiations and neutrinos) of blazars. Here, we consider a simple one-zone model with fewer parameters to explore the allowed parameter space by comparing the theoretical expectations with the observations of electromagnetic radiation and  high-energy neutrinos. Then, the~conclusions can help us to differentiate whether the one-zone model is valid or whether the two-zone model has to be~invoked.

For further simplification, we consider a spherical blob region in the jet where the accelerations and interactions of all electrons and protons take place. Such a scenario is the so-called one-zone lepto-hadronic model, which has been well developed to explain the SED and the neutrino emission of blazars. For~the case of TXS 0506+056,~\cite{cerruti2019leptohadronic} modeled the SED and neutrino flux in both the proton-synchrotron scenario and the $p\gamma $ scenario by~using the developed method for hadronic processes in blazars (see~\cite{kelner2008energy,bottcher2013leptonic}). Their solutions have degeneracy in parameters, especially in a big parameter space of magnetic field strength B 
 and blob radius $R$. Recently,~\cite{gasparyan2021time} developed a time-dependent code that follows the time evolution of the isotropic distribution functions of all particles~involved.

The parameters in most numerical calculations can vary by several orders of magnitude~\citep{bottcher2013leptonic,cerruti2015hadronic,zech2017expected,diltz2015time,jimenez2021katu,murase2018blazar,baring2019time}. This uncertainty may lead to quite different physical conditions of blazar jets and create difficulties for us to study the AGN environment. Here, we provide an analytical method to explore the viable parameter space and study the consequent constraints given by the observations. \cite{nalewajko2014constraining}  constrained the location of the blob and the bulk Lorentz factor through the electromagnetic observations of blazars in the framework of the leptonic process. However, the~observations of high-energy neutrinos from blazars could be another criterion to explore the model parameter space. In~this paper, we adopt a similar analytical method to constrain the radius of the blob and the bulk Lorentz factor in the framework of the lepto-hadronic one-zone model for BL Lacs by focusing on the $p\gamma $ scenario. We use a combination of constraints from the observed variability timescale of flare~$t_{\rm var}$, the~SSC luminosity~$L_{\rm SSC}$, the optical depth for gamma-ray photon~$\tau _{ \gamma \gamma}$, the~gamma-ray photon luminosity~$L_{ p\gamma}$ and the neutrino luminosity~$L_{\nu}$ produced by the hadronic process. Then, we apply our method to TXS 0506+056 and PKS 0735+178 for further studies. We emphasize that our method is based on the simple lepto-hadronic model without an external photon field (such as radiation from a broad line region or dust torus) and FSRQs have a non-negligible radiation component from BLR; hence, we only select BL Lacs as potential sources. The~cosmological parameters $H_0=69 \ \rm{km} \ \rm{s^{-1}} \ \rm{Mpc^{-1}}$, $ \Omega_{M}=0.286$ and~$\Omega_{\Lambda}=0.714$  \citep{bennett20141} are~applied.

This paper is organized as follows. We demonstrate our derivation in Section~\ref{derivation}. Then, we present the application to TXS 0506+056 in Section~\ref{section:TXS}. A further study on another BL Lacs, i.e.,~PKS 0735+178, is presented in Section~\ref{further study}. Our results and conclusions are discussed in Section~\ref{conclusion}.

\section{Derivation of~Constraints}
\label{derivation}

First, we set the rules for notation: Physical quantities measured in a co-moving frame will be  indicated by 
 a prime, those in~an AGN frame will be indicated by  a superscript ‘star’ (e.g.., $t'$ and $t^*$) and those in an observer frame will be indicated by  nothing. The peak luminosity of the SED for all kinds of radiation is $L_{\rm i}=\nu L_{i,\nu}$, in~contrast, the bolometric luminosity is presented as $L_{i,\rm bol}=\int{ L_{i, \nu}}\ d\nu$. We assume a spherical uniform emission region (blob) in a jet with radius $R'$, a propagating velocity $\beta =v/c$ (c is the speed of light) and a Lorentz factor $\Gamma =\left( 1-\beta ^2 \right) ^{-1/2}$. For~an observer on Earth, we should use the Doppler factor $\delta _{\rm D}=\left[ \Gamma \left( 1-\beta \cos \theta _{\rm obs} \right) \right] ^{-1}$ for transformations. The~relation between $\delta _{\rm D}$ and $\Gamma$ depends on the observer's viewing angle $\theta _{\rm obs}$ with respect to the blob propagating direction and opening angle $\theta$ of the blob, see more details in Section~2 of~\cite{nalewajko2014constraining}.


\subsection{Constraint from Variability~Timescale}
\label{time}
In order to avoid temporal integrations over different portions of the blob, the~timescale for variation in the radiation, $t'_{\rm var}$, should be longer than the light crossing time $t'_{\rm lc}\sim R'/c$~\citep{dermer2009high}, i.e.,
\begin{equation}
R'\lesssim c\cdot t'_{\rm var}=\frac{c\delta _{\rm D}t_{\rm var}}{\left( 1+z \right)}\ .
\end{equation}

Then,
\begin{equation}
\Gamma \gtrsim \frac{R'\left( 1+z \right)}{ct_{\rm var}}\cdot \left( \frac{\delta _{\rm D}}{\Gamma} \right) ^{-1}\ ,
\label{eq:time}
\end{equation}
where $t_{\rm var}$ is the variation timescale of the light curve, varying from days to months for different~sources.

\subsection{Constraint from SSC~Luminosity}

In a $p-\gamma $ scenario, gamma-ray photons from high-energy humps are mainly produced by the SSC process with a possible additional contribution from hadronic interactions~\citep{gao1807interpretation}; therefore, the~peak of the SSC spectrum should be less than the high-energy peak of the SED (which is in GeV, measured by Fermi-LAT), i.e.,
\begin{equation}
L_{\rm SSC}\leq L_{\gamma}\ .
\label{cons:SSC}
\end{equation}

We have the relation
\begin{equation}
\frac{L_{\rm SSC}}{L_{\rm syn}}\sim g_{\rm SSC}\left( \frac{u'_{\rm syn}}{u'_{\rm B}} \right)\ , 
\label{rela-ssc}
\end{equation}
because the input radiation field of the SSC process is the same photon field produced by electron-synchrotron radiation, where $u_{\rm B'}={B'^2}/{8\pi}$ and $u'_{ \rm syn}={L_{\rm syn}}/{(4\pi c\delta _{\rm D}^4{R'}^2)}$ are the energy densities of magnetic field and synchrotron radiation, respectively. $g_{\rm SSC}=\left( \frac{L_{ \rm SSC}}{L_{\rm syn}} \right) /\left( \frac{L_{\rm SSC,bol}}{L_{\rm syn,bol}} \right) \sim \frac{3}{4}$ is a bolometric correction factor (mainly due to the spectral shape and source geometry)~\citep{nalewajko2014constraining}. Furthermore, for electron-synchrotron radiation, we have the peak frequency $\nu _{\rm l,p}\sim 3\times 10^6 {\gamma^{'}}_{\rm e}^{2}B'\delta _{\rm D}/\left( 1+z \right) $, and~$\nu _{\rm h, p}\sim  {\gamma^{'}}_{\rm e}^{2}\nu _{\rm l, p}$ is the peak frequency of inverse Compton scattering. The combination of the two relations yields
\begin{equation}
B'\sim \frac{\left( 1+z \right)}{3\times 10^6\delta _{\rm D}}\cdot \left( \nu _{\rm l, p}^2/\nu _{\rm h, p} \right) \ .
\label{eq:B}
\end{equation}

Combining Equations~(\ref{rela-ssc}) and~(\ref{eq:B}), we obtain a constraint on $\Gamma$:
\begin{equation}
\Gamma \simeq \left[ \frac{3\times 10^6}{\left( 1+z \right)} \right]\left(\frac{\nu _{\rm h,p}L_{\rm syn}}{\nu _{\rm l, p}^2R'}\right)\left( \frac{2g_{\rm SSC}}{cL_{\rm SSC}} \right) ^{1/2}\left( \frac{\delta _{\rm D}}{\Gamma} \right) ^{-1}\ ,
\label{eq:ssc}
\end{equation}
with the constraint condition in Equation~(\ref{cons:SSC}).

\subsection{Constraint from Optical~Depth}
The high-energy photon may be absorbed by the low-energy photon field, leading to the maximum observed photon energy. The~peak cross-section for $\gamma\gamma$ annihilation is~$\sigma _{ \gamma \gamma}\sim \sigma _{\rm T}/5$, where $\sigma _{\rm T}$ is the Thomson cross-section. In the observer frame, we have a relation between the energy of the gamma-ray photon and the soft photon:
\begin{equation}
E_{\rm soft}\sim \frac{3.6\left( m_{\rm e}c^2 \right) ^2\delta _{\rm D}^2}{\left( 1+z \right) ^2E_{ \gamma}}.
\label{eq:ES}
\end{equation}

The optical depth for $\gamma\gamma$ annihilation is estimated as:
\begin{equation}
\tau _{\gamma \gamma}=\sigma _{ \gamma \gamma}n'_{\rm s}R'\sim \frac{\left( 1+z \right) \sigma _TL_{\rm soft}E_{\gamma}}{72\pi \left( m_{\rm e}c^2 \right) ^2c\delta _{\rm D}^5R'} .
\label{eq:depth}
\end{equation}

From the SED of TXS 0506+056~\citep{telescope2018multimessenger}, we can obtain the maximum gamma-ray energy of $E_{\max}\sim 5\times 10^{11}\,\rm eV$. To obtain $L_{\rm soft}$, we set a critical point~$E_{\rm X,0}$, which is the demarcation point between two humps in the SED. Then, we connect the soft photon luminosity and the critical point luminosity $L_{\rm X,0}$ via a spectral index $\alpha$ (indicating the slope of the SED, i.e.,~$\nu F_{\nu} \propto \nu^{\alpha}$), with~the value of $\alpha$ depending on the location of $E_{\rm soft}$ ($E_{\rm soft}>E_{\rm X,0}$ or $E_{\rm soft}<E_{\rm X,0}$), i.e.,
\begin{equation}
L_{\rm soft}=L_{\rm X,0}\left( \frac{E_{\rm soft}}{E_{\rm X,0}} \right) ^{\alpha}.
\label{eq:spec}
\end{equation}

Bringing this relation back to Equation~(\ref{eq:depth}), we obtain
\begin{equation}
\tau _{\gamma \gamma}=\frac{\sigma _{\rm T}3.6^{\alpha}\left( 1+z \right) ^{1-2 \alpha}E_{\max}^{1-\alpha}L_{\rm X,0}}{72\pi \left( m_{\rm e}c^2 \right) ^{2-2\alpha}c\delta _{\rm D}^{5-2 \alpha}E_{\rm X,0}^{\alpha}R'}.
\label{eq:depth index}
\end{equation}

The observations of gamma-ray photons with the maximum energy imply that \mbox{$\tau _{\gamma \gamma}\left( E_{\max} \right) \lesssim 1$}. Note that the gamma-ray photon energy with $\tau_{\gamma \gamma}\left( E_{\gamma} \right)\simeq 1$ may be around the cut-off position of the SED and smaller than the observed maximum gamma-ray energy $E_{\max}$; however, it will not affect our estimation significantly since the value of $\tau _{\gamma \gamma}\left( E_{\max} \right)$ is at most around a few. Then, we achieve our formula:
\begin{equation}
\Gamma \gtrsim \left[ \frac{\sigma _{\rm T}3.6^{\alpha}\left( 1+z \right) ^{1-2\alpha}E_{\max}^{1-\alpha}L_{\rm X,0}}{72\pi \left( m_{\rm e}c^2 \right) ^{2-2\alpha}cE_{\rm X,0}^{\alpha}R'} \right] ^{1/\left( 5-2\alpha \right)}\left( \frac{\delta _{\rm D}}{\Gamma} \right) ^{-1}.
\label{eq:cons-dep}
\end{equation}

We stress again that in the general case, $E_{\rm soft}$ could be larger or less than $E_{\rm X,0}$, leading to a different spectral index in Equation~(\ref{eq:spec}) and~thus a different constraint. The~value of $E_{\rm soft}$ is approximately estimated from Equation~(\ref{eq:ES}) with respect to different sources; we will briefly discuss this in the next~section.

\subsection{Constraint from the Hadronic~Process}
The $p\gamma $ process in the blob will produce gamma-ray photons and high-energy neutrinos through the photomeson process and  Bethe--Heitler (BH) pair production. Since the BH process will not produce neutrinos and contribute less to the final gamma-ray radiation, we will focus our study on the photomeson process. The~total produced gamma-ray photon luminosity ($L_{ p\gamma}$) and the neutrino luminosity ($L_{\nu}$) in the blob are calculated via:
\begin{equation}
L_{ p\gamma}=\frac{4}{3}\pi R'^{3}\delta _{\rm D}^{4}m_{\rm p}c^2\int_1^{\gamma' _{\rm p,\max}}{\frac{5}{8}f_{ p\gamma}\left( \gamma' _{\rm p} \right) \gamma' _{\rm p}}Q\left( \gamma' _{\rm p} \right) \ d\gamma' _{\rm p}
\label{eq:lph}
\end{equation}
and
\begin{equation}
L_{ \nu}=\frac{4}{3}\pi R'^{3}\delta _{\rm D}^{4}m_{\rm p}c^2\int_1^{\gamma' _{\rm p,\max}}{\frac{3}{8}f_{ p\gamma}\left( \gamma' _{\rm p} \right) \gamma' _{\rm p}}Q\left( \gamma' _{\rm p} \right) \ d\gamma' _{\rm p},
\label{eq:neu}
\end{equation}
where $Q\left( {\gamma^{'}}_{\rm p} \right)$ is the proton injection spectrum with the form of $Q\left( \gamma' _{\rm p} \right) =Q_0{\gamma^{'}}_{\rm p}^{-q}$, in which $q$ is the injection spectrum index, $\gamma' _{\rm p}$ is the Lorentz factor of protons in a co-moving frame and ${\gamma^{'}}_{\rm p,\max}$ is the maximum Lorentz factor. $f_{  p\gamma}\left( {\gamma^{'}}_{\rm p} \right)$ is the efficiency of the photomeson process which has a complicated integral formation~\citep{stecker1968effect}; however, we have a simple relation between the gamma-ray energy and the proton energy as long as they interact with the same soft photon field~\citep{xue2019minimum}:
\begin{equation}
E_{\rm p}\sim 3\times 10^5E_{\gamma}
\label{eq:re-pro}
\end{equation}
and relation between $f_{ p\gamma}\left( \gamma _{\rm p} \right)$ and $\tau _{\gamma \gamma}$:
\begin{equation}
f_{ p\gamma}\left( \gamma _{\rm p} \right) \sim 10^{-3}\tau _{\gamma \gamma} .
\end{equation}

Notice that both relations are valid in the observer frame. Combining the relations above and Equation~(\ref{eq:depth index}) together and changing the reference frame to a co-moving frame, we achieve a simplified formation for $f_{ p\gamma}\left( \gamma' _{\rm p} \right)$:
\begin{equation}
f_{ p\gamma} \sim 10^{-3}\cdot \left( 3\times 10^5 \right) ^{\alpha -1}\cdot\\
\frac{\sigma _{\rm T}3.6^{\alpha}\left( 1+z \right) ^{-\alpha}\left( m_{\rm p}c^2 \right) ^{1-\alpha}L_{\rm X,0}{\gamma^{'}}_{\rm p}^{\left( 1-\alpha \right)}}{72\pi \left( m_{\rm e}c^2 \right) ^{2-2\alpha}c\delta _{\rm D}^{4-\alpha}E_{\rm X,0}^{\alpha}R'} 
\label{eq:eff}
\end{equation}

Furthermore, we have the proton luminosity $L_{\rm p}^{*}$ in the AGN frame, which is estimated as the lowest power for the jet:
\begin{equation}
L_{\rm p}^{*}=\pi R'^{3}m_{\rm p}c^2\Gamma ^2\int_1^{\gamma' _{\rm p,max}}{\gamma' _{\rm p}}Q\left( \gamma' _{\rm p} \right) \ d\gamma' _{\rm p}.
\label{eq:lp}
\end{equation}

With Equations~(\ref{eq:lph}), (\ref{eq:eff}) and~(\ref{eq:lp}),~functions for $L_{p\gamma}$ and $R'$ are obtained:
\begin{equation}
\begin{split}
\Gamma ^{2-\alpha}\sim \frac{4}{3}\cdot \frac{5}{8}\cdot 10^{-3}\cdot \left( 3\times 10^5 \right) ^{\alpha -1}\cdot \frac{2-q}{3-\alpha -q}\cdot\\
\boldmath{\frac{{\gamma^{'}}_{\rm p,\max}^{3-\alpha -q}-1}{{\gamma^{'}}_{\rm p,\max}^{2-q}-1}\cdot\frac{L_{\rm p}^{*}}{L_{p\gamma}}\cdot\left( \frac{\delta _{\rm D}}{\Gamma} \right) ^{\alpha}\cdot B\left( R' \right)},
\end{split}
\label{eq:cons-hard}
\end{equation} 
where $B\left( R' \right) =\frac{\sigma _{\rm T}3.6^{\alpha}\left( 1+z \right) ^{-\alpha}\left( m_{\rm p}c^2 \right) ^{1-\alpha}L_{\rm X,0}}{72\pi \left( m_{\rm e}c^2 \right) ^{2-2\alpha}cE_{\rm X,0}^{\alpha}R'}$. For~neutrino luminosity $L_{ \nu}$, the~formula is similar but  the prefactor $\frac{5}{8}$ is replaced  with $\frac{3}{8}$ and $L_{ p\gamma}$ with $L_{ \nu}$.  

Due to the possibility of other contributions (e.g., the~SSC and the BH process), the~constraints can be obtained by the fact that the observed peak luminosity ($L_{\gamma}$) of Gev gamma-ray band by Fermi-LAT should be larger than the contributions of the photomeson process which would be generally cascaded peaking in the GeV~band.


For the high-energy neutrino, the~expected luminosity from the photomeson process should be larger than the actual observed value or conservatively larger than 0.003 times the observed value to ensure the detection of the high-energy neutrino is at the $3\sigma$ significance level. Note that $L_{\nu}$ in Equation~(\ref{eq:neu}) is the luminosity for all-flavor neutrinos, while many observed values in the literature are only for muon neutrinos and anti-muon neutrinos. In~conclusion, we have the constraints:
\begin{equation}
L_{ p\gamma}<L_{\gamma}=4\pi D_{\rm L}^{2}F_{\gamma},
\label{pggamma}
\end{equation}
\begin{equation}
L_{\nu}>4\pi D_{\rm L}^{2}F_{\nu},
\label{pgneutrino}
\end{equation}
or more conservatively using
\begin{equation}
L_{\nu}>4\pi D_{\rm L}^{2}F_{\nu}\times 0.003,
\label{pgneutrino003}
\end{equation}
where $F_{\nu}$ is the flux of all-flavor neutrinos based on  one high-energy neutrino detection during the corresponding time window. Then, combining these with Equation~(\ref{eq:cons-hard}) by replacing $L_{p\gamma}$ ($L_{\nu}$) therein, one can give constraints for the hadronic~process.

\section{Application to TXS 0506+056}
\label{section:TXS}
In this section, we will apply the above-derived constraints to the specific neutrino source TXS 0506+056, which coincides with the IC170922A event. Observational values can be obtained from the SED shown in~\cite{telescope2018multimessenger} and summarized in Table~\ref{table:parameters}. In~addition, we assume $\delta _{\rm D}/\Gamma\sim1$ for all constraints. Here, since the photon field presents two different spectral indexes, i.e.,~$\alpha \sim-0.48$ and $0.31$ (here, $\nu F_{\nu} \propto \nu^{\alpha}$), below~and above the critical photon energy $E_{\rm X,0}$, respectively, for the TXS 0506+056 observations, we evaluate the energy of soft photon field by whether they can participate in the $\gamma\gamma$ annihilation (for the constraint from optical depth) and the photomeson process (for the constraint from the hadronic~process). 

For  $\gamma\gamma$ annihilation, the~maximum gamma-ray energy and critical photon energy are estimated as $E_{\max}\sim 5\times 10^{11}\,\rm eV$ and $E_{\rm X,0}\sim 4\times 10^{3}\,\rm eV$. One has the typical energy $E_{\rm soft}\sim 1\times\delta _{\rm D}^{2}\,\rm eV$ from Equation~(\ref{eq:ES}) to attenuate the gamma-rays with maximum energies. Hence, $E_{\rm soft}<E_{X,0}$ when $\delta _{\rm D}\lesssim 63$ and $E_{\rm soft}$ is located in the soft X-ray band and the spectral index is adopted as $\sim$$-$0.48.

\begin{table}[h] 
\caption{\label{table:parameters} Values for the parameters of TXS 0506+056 and~PKS 0735+178. Data are from~\cite{telescope2018multimessenger,sahakyan2022multi}, while $\gamma' _{\rm p,\max}$, $\gamma' _{\rm p,\min}$, $q$ and $L_{\rm p}^{*}$ are free~parameters.}
\centering
\begin{tabular}{@{}llll@{}}
\hline
			{\it \rm \textbf{Parameter}} & {\it \rm \textbf{Symbol}} & {\it \rm \textbf{TXS 0506+056}}& {\it \rm \textbf{PKS 0735+178}}\\
			\midrule
			Syn-radiation peak flux & $F_{\rm syn,p}$ & { $4\times 10^{-11} \,\rm erg/s/cm^{2}$} & { $3.9\times 10^{-11} \,\rm erg/s/cm^{2}$}\\
			gamma-ray peak flux & $F_{ \gamma,p}$&{ $5\times 10^{-11}\,\rm erg/s/cm^{2}$}&{ $8.6\times 10^{-11}\,\rm erg/s/cm^{2}$}\\
			Neutrino flux& $F_{\nu}$&{ $5.4\times 10^{-10} \,\rm erg/s/cm^{2}$}&{ $9.4\times 10^{-11} \,\rm erg/s/cm^{2}$}\\
			Critical photon flux&  $F_{\rm X,0} $& { $8\times 10^{-13}\,\rm erg/s/cm^{2}$}& { $9.3\times 10^{-13}\,\rm erg/s/cm^{2}$}\\
			Critical photon energy & $ E_{\rm X,0} $&{ $4.0\times10^{3}$\,\rm eV}&{ $5.8\times10^{3}$\,\rm eV}\\
			Maximum photon energy& $E_{\rm max}$ &{ $5\times10^{11}$\,\rm eV}&{ $3.4\times10^{9}$\,\rm eV}\\
			Low energy peak frequency& $ \nu _{\rm l,p} $& $ 1\times10^{15}\,\rm Hz$& $ 1.3\times10^{15}\,\rm Hz$\\
			High energy peak frequency& $ \nu _{\rm h,p} $& $ 5\times10^{22}\,\rm Hz$& $ 4.5\times10^{22}\,\rm Hz$\\
			Redshift& $ z$& $0.3365 $& $0.65 $\\
			Spectral index\footnote{
The spectral index ($\nu F_{\nu} \propto \nu^{\alpha}$) could be different below $E_{\rm X,0}$ and above $E_{\rm X,0}$. For~TXS0506+056, two different indexes are involved, while for PKS 0735+178, only one is involved. See text for details.}  & $ \alpha $&$-0.48 \,\rm{or} \,0.31 $& $0.44 $\\
			Time variation & $ t_{\rm var} $& $  $\,\rm one week& $ 5000 $\,\rm~s\\
			Maximum Lorentz factor&$\gamma' _{\rm p,\max}$&$10^6$&$10^6$\\
			Minimum Lorentz factor&$\gamma' _{\rm p,\min}$&$1$&$1$\\
			Injection index&$q$ & $1.8-2.2 $& $1.8-2.2 $\\
			Proton luminosity \mbox{(AGN frame)}&$L_{\rm p}^{*}$&$10  L_{\rm Edd}-10^3 L_{\rm Edd}$&$10  L_{\rm Edd}-2\times10^3  L_{\rm Edd}$\\
			\bottomrule
\end{tabular}

\end{table}

As the results demonstrate below, since for $\delta _{\rm D} \gtrsim 63$, the~constraints obtained by the optical depth of maximum energy photons become negligible so that it will not affect the final parameter space, we achieve constraints through the opacity of maximum energy photons considering the soft photon field with the spectral index of $\alpha\sim -0.48$ only. For~the photomeson process, $E_{\rm soft}$ depends on $\gamma'_{\rm p}$ via Equations~(\ref{eq:ES}) and (\ref{eq:re-pro}). To~obtain a observational neutrino energy with tje range of 1 TeV--1 PeV, $\gamma _{\rm p}\sim \delta_{\rm D} \gamma'_{\rm p}$ should be in the range of $10^3-10^6$, indicating that the soft photon energy that exceeds the threshold of photomeson process is $E_{\rm soft} \gtrsim 10^4 (\gamma_{\rm p}/10^6)^{-1}(\delta_{\rm D}/10)^2\,\rm eV$. For~$\delta_{\rm D}$ slightly larger than a few, all concerned protons will interact with the soft photons with energies above $E_{\rm X,0}$, so constraints from the hadronic process can be reached by considering the soft photon field with a spectral index of $\alpha\sim 0.31$ only. Other parameters needed are shown in Table~\ref{table:parameters}, where we have multiplied (anti-)muon neutrino flux from~\cite{aartsen2017icecube} by 3 to obtain all flavor fluxes. We have four free parameters: $\gamma' _{\rm p,\max}$, $\gamma' _{\rm  p,\min}$, $q$ and $L_{\rm p}^{*}$. Here, we take $\gamma' _{\rm p,\min}=1$ and $\gamma' _{\rm p,\max}=10^{6}$ to normalize the proton luminosity. In~addition, such a range of $\gamma'_{\rm p}$ can achieve the optimistic neutrino production around PeV, yielding relatively conservative constraints from the hadronic process. In addition, we keep $q$ and $L_{\rm p}^{*}$ free to explore their influence on the constraints of the parameter~space.


 Among all situations concerned, we find that the injected proton luminosity should exceed $2.5\times10^{49}\,\rm erg/s$, about $42 L_{\rm Edd}$, while the Eddington luminosity for this source is estimated as $\sim 10^{47.8}\, \rm erg/s$ by assuming the mass of the central black hole is $5\times10^9\,\rm M_{\odot}$, since the mass of this source is uncertain. The~injected luminosity for a proton must be larger than $2.5\times10^{49} \,\rm erg/s$ for $q= 1.8$, $5.0\times10^{49} \,\rm erg/s$ for $q= 2.0$ and $2.0\times10^{50}\,\rm erg/s$ for $q= 2.2$, otherwise there is no allowed parameter space. The required lower limits of $L_{\rm p}^{*}$ in our results are higher than in~\cite{xue2019minimum}, which may be caused by the different $\gamma' _{\rm p,\max}$ selection. We note that with a higher $\gamma' _{\rm p,\max}$, the~allowed value for $\Gamma$ will increase. However, this will overestimate the neutrino luminosity. A~sample result for the constraints and  parameter space is demonstrated in Figure~\ref{Fig:Lp=e2ledd,q=2} under~the conditions of $q= 2.0, L_{\rm p}^{*}= 6\times10^{50}\,{\rm erg/s}\simeq 10^{3}L_{\rm Edd}$, where the allowable parameter space is~highlighted. 
 
\begin{figure}[h]
\vspace{-12pt}
	\centering
	\includegraphics[width=0.6\textwidth]{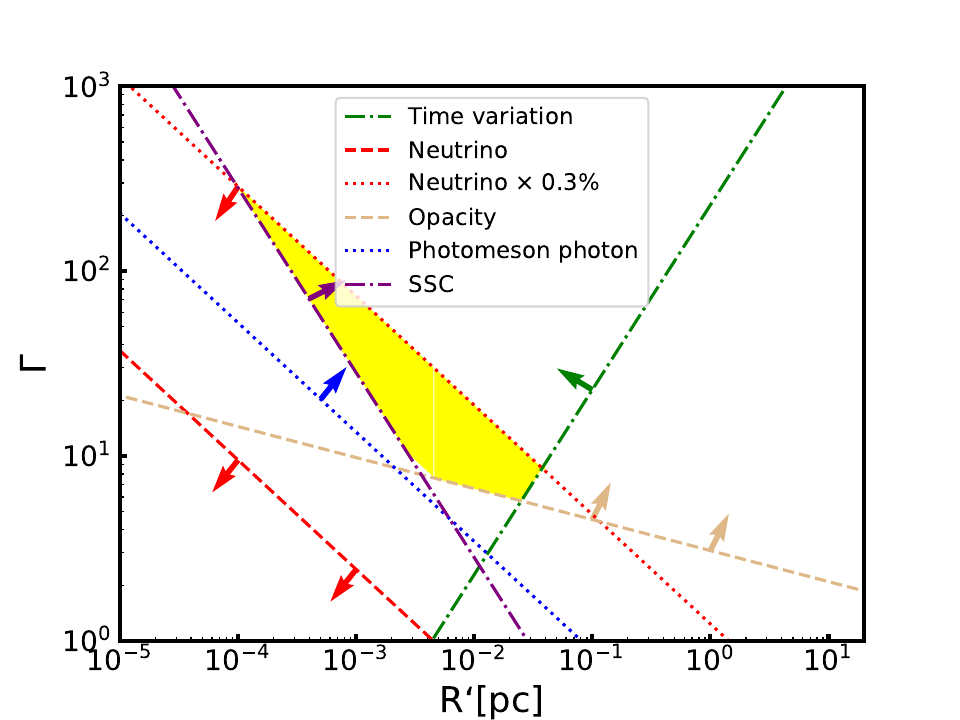}
	\caption{One situation of constraints for $\Gamma$ and $R'$, where $q=2.0$ and $L_{\rm p}^{*}= 10^{3}L_{\rm Edd}$. The highlighted region shows the available parameter space. The arrows indicate the allowed parameter space for diverse constraint methods.The~green dot-dashed line is from Equation~(\ref{eq:time}); the purple dot-dashed line is from Equation~(\ref{cons:SSC}); the yellow dot-dashed line is from Equation~(\ref{eq:cons-dep}); the blue and red lines are all from Equation~(\ref{eq:cons-hard}), where the neutrino fluxes of the red dashed line and the red dotted line  are $F_{\nu}\times0.3\%$ (Equation~(\ref{pgneutrino003})) and $F_{\nu}$ (Equation~(\ref{pgneutrino})), respectively. The allowed parameter values, in~this case, are 5.9--299.9 for $\Gamma$ and  $1.0\times10^{-4}$--$3.6\times10^{-2}$ pc for $R'$.}
	\label{Fig:Lp=e2ledd,q=2}
\end{figure}

\begin{figure}[h]

\includegraphics[width=0.47\linewidth]{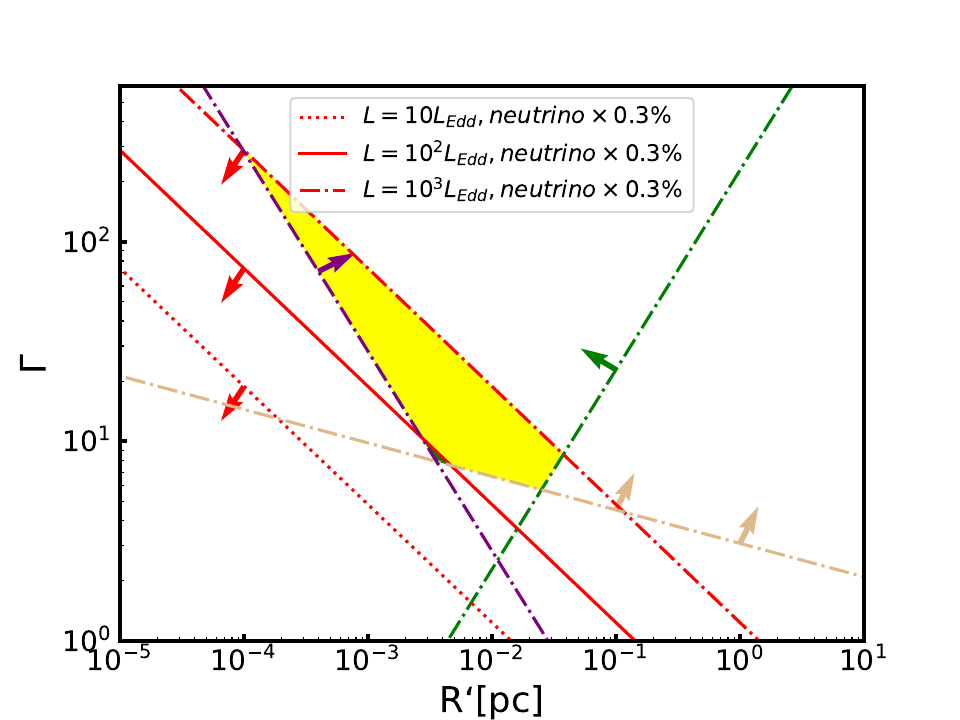}
\includegraphics[width=0.49\linewidth]{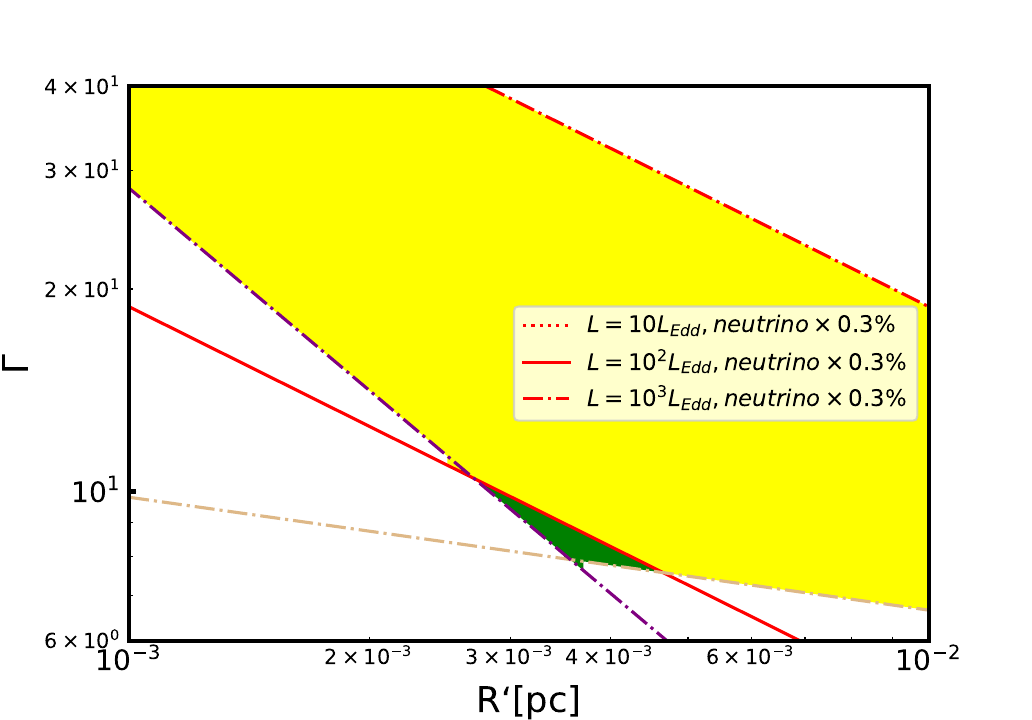}
\caption{(\textbf{Left panel}) Different 
 constraints when q is fixed as 2.0, while $L_{\rm p}^{*}= 10 L_{\rm Edd}, 10^2 L_{\rm Edd}$ and $10^3 L_{\rm Edd}$, respectively. Constraints of time variability, SSC and~opacity will not vary with q and $L_{\rm p}^{*}$; hence, we only demonstrate the constraints given by neutrinos with $0.3\%$ of the observed flaring neutrino flux (dashed, solid and dot-dashed line in red) for different $L_{\rm p}^{*}$ values, while ignoring the constraints of gamma-rays from the photomeson process and neutrinos with $100\%$ of the observed flaring neutrino flux, since they have no effects on the parameter space in a selected condition. The arrows indicate the allowed parameter space for diverse constraint methods as the same as in Fig.1.There are allowed parameter spaces in the condition of $L_{\rm p}^{*}= 10^2 L_{\rm Edd}$  (highlighted region in green) and $L_{\rm p}^{*}= 10^3 L_{\rm Edd}$ (highlighted region in yellow and green both), suggesting an ultra-high proton injection luminosity is required. The right panel is zoomed in to visually distinguish the green region.}
\label{Fig:qfixed}

\end{figure}

With a fixed q, the~area of allowed parameter space  increases with the injected proton luminosity, as~shown in Figure~\ref{Fig:qfixed}, where $L_{\rm p}^{*}= 10^3L_{\rm Edd}$, $10^2 L_{\rm Edd}$ and $10 L_{\rm Edd}$, respectively. The constraints of time variability, SSC and~opacity will not vary with q and $L_{\rm p}^{*}$; hence, we only demonstrate the constraints given by a neutrino flux larger than $0.3\%$ of the observed flaring neutrino flux (Equation~(\ref{pgneutrino003})) for different $L_{\rm p}^{*}$ values (dashed, solid and dot-dashed line in red) and ignore the inconsequential constraints of gamma-rays from the photomeson process (Equation~(\ref{pggamma})) and neutrino flux is the same as the observed flaring neutrino flux (Equation~(\ref{pgneutrino})). The~constraint of gamma-rays from the photomeson process could affect the parameter space only in extreme conditions: the injected luminosity for a proton must be larger than \mbox{$4.0\times10^{50} \,\rm erg/s$} for $q= 1.8$, $9.0\times10^{50} \,\rm erg/s$ for $q= 2.0$ and $4.0\times10^{51}\,\rm erg/s$ for $q= 2.2$, which are not used here. Hereafter, we take $L_{\rm p}^{*}= 10^2L_{\rm Edd}$, $q= 2.0$ as a benchmark for comparison. From~Figure~\ref{Fig:qfixed}, one may find that the area of allowed parameter space varies dramatically by changing $L_{\rm p}^{*}$ and the available values for $\Gamma$ vary from $\sim$10 to $\sim$100. 
We summarize the allowed value range in Table~\ref{tab:totalava}. Note that the values summarized are only upper and lower limits for $\Gamma$ and $R'$.

\begin{table}[h]
	\caption{Available values for $R'$ and $\Gamma$ in various conditions, where "---" means no allowable parameter space in such situations, for the case of TXS 0506+056  }
	\label{tab:totalava}
	\renewcommand\tabcolsep{2.0pt}
	\begin{tabular}{ccccccc}
		\toprule
		\multirow{2}{*}{\textbf{Conditions}}&\multicolumn{2}{c}{\textbf{q=1.8}} &\multicolumn{2}{c}{\textbf{q=2.0}}&\multicolumn{2}{c}{\textbf{q=2.2}}\\
		\cmidrule(r){2-3}\cmidrule(r){4-5}\cmidrule(r){6-7}
	&  \boldmath{$R'$}\textbf{[pc]} &  \boldmath{$\Gamma$} 
		&  \boldmath{$R'$}\textbf{[pc]} &  \boldmath{$\Gamma$}  
		&  \boldmath{$R'$}\textbf{[pc]} &  \boldmath{$\Gamma$}\\
		\midrule
		$L_{\rm p}^{*}=10L_{\rm Edd}$ & --- & --- &--- & --- & ---&---\\
		$L_{\rm p}^{*}=50L_{\rm Edd}$& $2.2\times10^{-3}-6.1\times10^{-3}$& $7.2-12.3$ &--- &--- &--- &---\\
		$L_{\rm p}^{*}=10^2L_{\rm Edd}$ & $7.7\times10^{-4}-1.6\times10^{-2}$&$6.3-37.3$ &$2.3\times10^{-3}-4.7\times10^{-3}$ &$7.5-10.5$ &---&---\\
		$L_{\rm p}^{*}=10^3L_{\rm Edd}$ & $2.7\times10^{-5}-5.0\times10^{-2}$&$5.9-1023.4$ &$1.0\times10^{-4}-3.6\times10^{-2}$ &$5.9-299.9$ &$6.2\times10^{-4}-2.0\times10^{-2}$ &$6.1-44.9$\\
		\bottomrule
	\end{tabular}
\end{table}

\begin{table}[h]
	\caption{Available values for $R'$ and $\Gamma$ in various conditions, where "---" means no allowable parameter space in such situations, for the case of PKS 0735+178  }
	\label{table:result2}
	\renewcommand\tabcolsep{1pt}
	\begin{tabular}{ccccccc}
		\toprule
		\multirow{2}{*}{\textbf{Conditions}}&\multicolumn{2}{c}{\textbf{q=1.8}} &\multicolumn{2}{c}{\textbf{q=2.0}}&\multicolumn{2}{c}{\textbf{q=2.2}}\\
		\cmidrule(r){2-3}\cmidrule(r){4-5}\cmidrule(r){6-7}
	&  \boldmath{$R'$}\textbf{[pc]} &  \boldmath{$\Gamma$} 
		&  \boldmath{$R'$}\textbf{[pc]} &  \boldmath{$\Gamma$}  
		&  \boldmath{$R'$}\textbf{[pc]} &  \boldmath{$\Gamma$}\\
		\midrule
		$L_{\rm p}^{*}=10L_{\rm Edd}$ & --- & --- &--- & --- & ---&---\\
		$L_{\rm p}^{*}=200L_{\rm Edd}$& $7.3\times10^{-4}-7.7\times10^{-4}$& $26.0-27.1$ &--- &--- &--- &---\\
		$L_{\rm p}^{*}=500L_{\rm Edd}$ & $1.4\times10^{-4}-1.1\times10^{-3}$&$26.0-138.7$ &$6.3\times10^{-4}-8.0\times10^{-4}$ &$26.0-31.7$ &---&---\\
		$L_{\rm p}^{*}=10^3L_{\rm Edd}$ & $4.3\times10^{-5}-1.4\times10^{-3}$&$26.0-463.0$ &$1.9\times10^{-4}-1.0\times10^{-3}$ &$26.0-108.5$ &---&---\\
		$L_{\rm p}^{*}=2\times10^3L_{\rm Edd}$ & $1.2\times10^{-5}-1.9\times10^{-3}$&$26.0-1558.1$ &$5.0\times10^{-5}-1.4\times10^{-3}$ &$26.0-392.5$ &$4.5\times10^{-4}-8.0\times10^{-4}$ &$26.0-44.3$\\
		\bottomrule
	\end{tabular}
\end{table}

\begin{figure}[h]

\vspace{-12pt}
\includegraphics[width=0.47\linewidth]{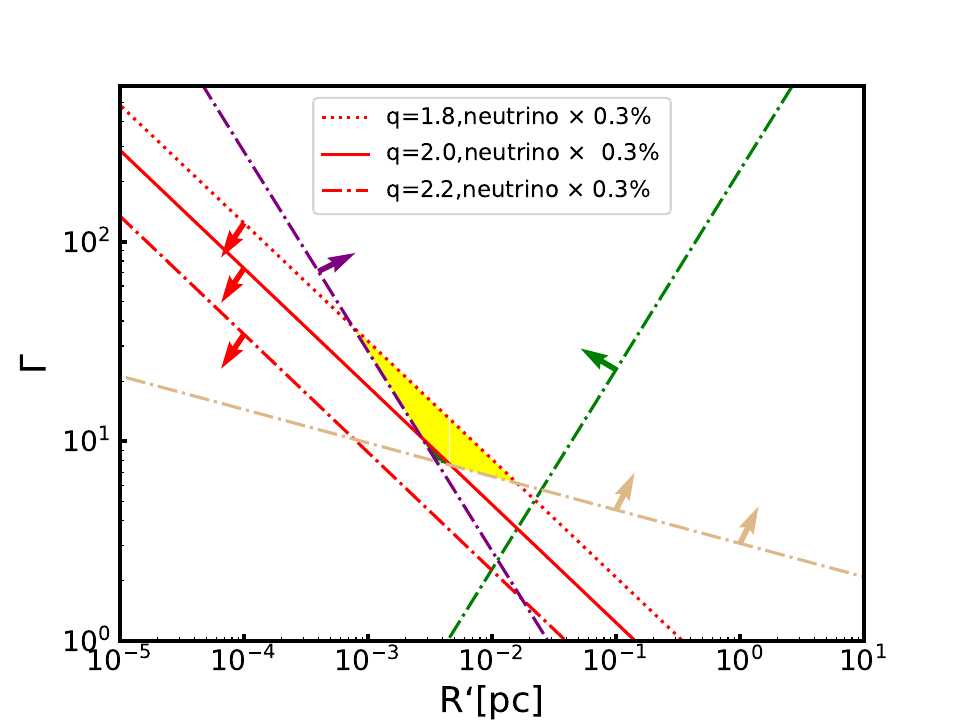}
\includegraphics[width=0.5\linewidth]{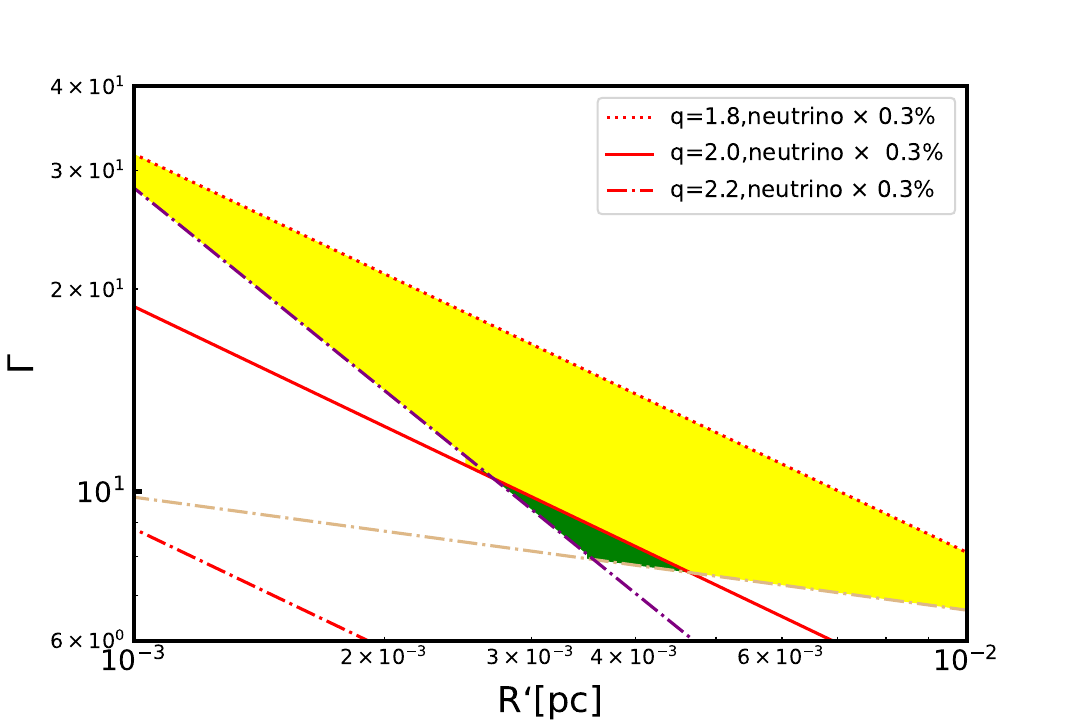}
\caption{A similar demonstration to Figure~\ref{Fig:qfixed}, except~  $L_{\rm p}^{*}= 10^2 L_{\rm Edd}$ is fixed while q is varied. There are allowed parameter spaces in the condition of {$q= 2.0$} (highlighted region in green) and {$q= 1.8$} (highlighted region in green and yellow in both).}
\label{Fig:Lpfixed}

\end{figure}

When $L_{\rm p}^{*}$ is fixed, the~area of the allowed parameter space decreases through increasing q, as~shown in Figure~\ref{Fig:Lpfixed}. Similar to Figure~\ref{Fig:qfixed}, we consider three different values of $q$, i.e.,~1.8, 2.0 and~2.2, and~a fixed proton luminosity of $L_{\rm p}^{*}= 10^2L_{\rm Edd}$. Comparing the two figures above, we found that the variation of the parameter is less sensitive to $q$ than to  $L_{\rm p}^{*}$. 
	

\section{Application to PKS 0735+178}
\label{further study}
For further study, we also investigated another BL Lac possibly associated with a neutrino event: PKS 0735+178 and IC-211208A. IC-211208A is a muon track event with an estimated energy of 172 TeV with a large statistical 90$\%$ localization error region of $\sim$13 square degrees. The~potential source PKS 0735+178 (z$\sim$0.65), an~intermediate synchrotron peaked BL Lac (ISP), is located  slightly outside of the error region of IC-211208A~\citep{sahakyan2022multi}. However, it was an outburst in $\gamma$ rays, X-rays and optical-UV at the time of the neutrino alert. Moreover, it might be associated with  three other neutrino events detected by Baikal-GVD~\citep{dzhilkibaev2021baikal}, the Baksan Underground Scintillation Telescope~\citep{petkov2021baksan} and Km3NeT undersea neutrino detectors~\citep{filippini2022search}. To~be consistent with the analysis of TXS 0506+056, we still used IceCube data to study this~source.

Similar to the case of TXS 0506+056, we first confirm some key values in our method. For~the neutrino flux $ F_{\nu}$, we estimate the effective area of the neutrino detector for IC-211208A as A$\sim$160 $\rm m^2$ using Figure~5 from~\cite{aartsen2020time} and the neutrino energy as $\sim$172 TeV. The duration time for neutrino emission is taken as the multiwavelength flare duration time of 3 weeks~\citep{sahakyan2022multi}. With~this duration time and effective area, the~neutrino flux can be calculated as  \mbox{$F_{\nu}$$\sim$$9.4\times10^{-11} \rm erg/cm^2/s$}. For~the spectral index $\alpha$, we first estimate the critical photon energy as $E_{\rm X,0}$$\sim$$5.8\times 10^3 \,\rm eV$ and the maximum photon energy as $E_{\rm max}$$\sim$$3.4\times 10^9 \,\rm eV$ from the SED of PKS 0735+178~\citep{sahakyan2022multi}. From \mbox{Equation~(\ref{eq:ES})}, we have $E_{\rm soft}\sim 100 \delta _{\rm D}^{2}\,\rm eV$ and for any $\delta _{\rm D} \gtrsim 7.5$, $E_{\rm soft}$ will be larger than the critical energy $E_{X,0}$. Furthermore, in this case, the constraints from time variation and SSC  already imply that $\delta _{\rm D}$ must be larger than 26.0, so the condition $\delta _{\rm D} \gtrsim 7.5$ is always satisfied and we can estimate the spectral index $\alpha$ for Equation~(\ref{eq:depth index}) as $\sim$ 0.44. The~spectral index for Equation~(\ref{eq:cons-hard}) can be obtained in a similar way. The  adopted parameters and the results are summarized in \mbox{Tables~\ref{table:parameters} and~\ref{table:result2}}.

Similar to the results of TXS 0506+056, a~super-Eddington luminosity for proton power is required ($M_{\rm SMBH}$$\sim$$6.3\times10^8 M_{\odot }$). There is an allowed parameter space only when $L_{\rm p}^{*}$ is larger than $200L_{\rm Edd}$. When $L_{\rm p}^{*}$ exceeds $2\times10^3L_{\rm Edd}$, there is an allowed parameter space under all conditions. Our results are consistent with the parameters chosen in~\cite{sahakyan2022multi}.

\section{Conclusions and~Discussion}
\label{conclusion}
In this work, we study the parameter space of the radius and bulk Lorentz factor of blobs with an analytical method in the framework of the lepto-hadronic one-zone model for BL Lacs. We use a combination of constraints from the observed variability timescale~$t_{\rm var}$, synchrotron self-Compton (SSC) luminosity~$L_{\rm SSC}$, optical depth for gamma-rays~$\tau _{\gamma \gamma}$, photon luminosity~$L_{ p\gamma}$ and neutrino luminosity~$L_{\nu}$ in the hadronic process. We apply our method to TXS 0506+056 and PKS 0735+178, then explore the allowed parameter space. We find that the allowed values for $\Gamma$ and $R'$ vary with different injected proton powers $L_{\rm p}^{*}$ and injection indexes q, and~are more sensitive to $L_{\rm p}^{*}$.
For two studied BL Lac--neutrino associations, a~proton luminosity significantly exceeding the Eddington luminosity is required to have an allowed parameter space for the simple lepto-hadronic one-zone~model. 



Our analytical constraints on the allowed parameter space based on the lepto-hadronic one-zone model should be more conservative than from the detailed numerical fitting to the electromagnetic radiation and neutrino spectrum of BL Lacs, since only some key spectral characteristics are selected to limit the parameter space. However, our conservative constraints have introduced a quite large proton luminosity compared to the Eddington luminosity, probably suggesting that the actual condition may disfavor the simple  one-zone lepto-hadronic model. As~a result, the~more complicated model may be invoked to explain the BL Lac--neutrino association event. Future multi-messenger observations could help us to determine the actual physical condition~further. 

In addition to the selected BL Lac--neutrino associations in this paper, some other possible BL Lac--neutrino associations have been reported as well, e.g.,~IC-200107A associated with 4FGL J0955.1+3551 \citep{2020ApJ...902...29P,2020A&A...640L...4G} and IC-141209A with GB6 J1040+0617~\citep{2019ApJ...880..103G}. However, the~data on their broadband electromagnetic radiation are inadequate to provide effective constraints. Besides,~potential associations between high-energy neutrino events and FSRQs, such as IC-35 and PKS B1424-418~\citep{kadler2016coincidence}, IC-190730A and PKS 1502+106~\citep{franckowiak2020patterns}, have been also reported. For~FSRQs, the~extra external photon field, especially Broad Line Region (BLR) radiation, could  serve as the seed photons of the EC process contributing to the observed gamma-rays and the $p\gamma$ process contributing to the high-energy neutrino radiation. The~latter constraint on the proton luminosity is more stringent and could be alleviated to allow a smaller proton luminosity to a certain~extent.


\bibliography{reference}{}
\bibliographystyle{aasjournal}



\end{document}